\documentclass[12pt]{article}
\usepackage{graphicx}

\newsavebox{\sboxpubnumber}
\newsavebox{\sboxpubdate}
\newcommand{\pubdate}[1]{\begin{lrbox}{\sboxpubdate}{#1}\end{lrbox}}
\newcommand{\pubnumber}[1]{\begin{lrbox}{\sboxpubnumber}{\begin{tabular}{l} #1 \\
				 \usebox{\sboxpubdate}
				 \end{tabular}}
                           \end{lrbox}
                           \pubblock}
\newcommand{\Title}[1]{\begin{center} {\Large #1 } \end{center}}
\newcommand{\Author}[1]{\begin{center}{ \sc #1} \end{center}}
\newcommand{\Address}[1]{\begin{center}{ \it #1} \end{center}}

\newcommand{\pubblock}{\rightline{
			\usebox{\sboxpubnumber}}}
\newenvironment{Abstract}{\begin{quotation}  }{\end{quotation}}
\newenvironment{Presented}{\begin{quotation} \begin{center}
             PRESENTED AT\end{center}\bigskip
      \begin{center}\begin{large}}{\end{large}\end{center}
      \end{quotation}}

\begin{document}

\begin{titlepage}
\pubdate{\today}                    
\pubnumber{SHEP 01-30} 

\vfill
\Title{Supersymmetric Inflation, Matter and Dark Energy}
\vfill
\Author{S. F. King}
\Address{Department of Physics and Astronomy,\\
University of Southampton, \\Southampton, SO17 1BJ, U.K.}
\vfill
\begin{Abstract}
In this talk, based on \cite{Kane:2001rb}, we emphasise that 
intermediate scale supersymmetric inflation
models are particularly attractive since
inflation, baryogenesis and the
relic abundance of cold dark matter (CDM) are all related by a set
of parameters which also affect particle physics collider
phenomena, neutrino masses and the strong CP problem. 
We also point out that the present day relic abundances 
of different forms of matter are (in principle) calculable
from the supersymmetric inflation model together with
a measurement of the CMB temperature and the Hubble constant.
From these relic abundances one can deduce the amount of the present day
dark energy (DE) density.

\end{Abstract}
\vfill
\begin{Presented}
    COSMO-01 \\
    Rovaniemi, Finland, \\
    August 29 -- September 4, 2001
\end{Presented}
\vfill
\end{titlepage}
\def\thefootnote{\fnsymbol{footnote}}
\setcounter{footnote}{0}

\section{Introduction}

The introductory parts of the talk
included the motivation for supersymmetry and inflation, 
and ten challenges facing a supersymmetric inflationary model.
This was followed by a detailed review of a particular intermediate
scale supersymmetric inflation model, as an example of a model which attempts
to meet all the challenges. These parts of my talk
can be found in the 
recent review article \cite{King:2001mt} to which we refer those
readers requiring more background information.
Here we state the omega problem in supersymmetric inflation,
give the motivation for intermediate scale inflation and introduce
a particular model. We then show how the parameters of this model
simultaneously control different phenomena in particle physics and
cosmology. Finally we show how the matter relic densities may be calculated
from such a model together with
a measurement of the CMB temperature and the Hubble constant,
and hence how the DE density may be deduced.

\section{The Omega Problem in Supersymmetric Inflation}

Within the framework of intermediate scale supersymmetric inflation
one may expect on general grounds
that our Universe contains sizeable relic abundances of baryons
(from e.g. leptogenesis), axions (a) (from the Peccei-Quinn (PQ)
solution to the strong CP problem), 
as well as weakly interacting massive particles
(WIMPS). In R-parity conserving SUSY the WIMP is identified
as the lightest supersymmetric particle (LSP), and this is
often assumed to be
the lightest neutralino of the minimal supersymmetric standard model
(MSSM). We shall argue from this perspective 
that the LSP could equally well be a lighter stable singlet
(singlino) identified with an axino or inflatino.

In general the ratio of the total density of the universe $\rho_{tot}$
to critical density $\rho_{crit}$
is given by $\Omega_{tot}$ where
\begin{equation}
\Omega_{tot} = \Omega_{\gamma}+\Omega_{matter}+ \Omega_{DE} 
\label{omega}
\end{equation}
and $\Omega_{\gamma}$, $\Omega_{matter}$, $\Omega_{DE}$ are the ratios of
radiation density $\rho_{\gamma}$, 
matter density $\rho_{matter}$
and DE density $\rho_{DE}$ to critical
density $\rho_{crit}$, and the radiation density is unimportant 
$\rho_{\gamma}\ll \rho_{matter}$.
Note that the critical density is a function of time 
and in the present epoch
$\rho_{crit}=3M_P^2H_0^2 =(3h^{1/2}\times 10^{-3}\ {\rm eV})^4\sim
(M_W^2/M_P)^4$ where $M_W$ is the weak scale, $M_P$ is the Planck
scale, and $H_0=100h {\rm km.s^{-1}Mpc^{-1}}$ 
is the present day Hubble constant, with $h=0.7\pm 10\%$.
From observation $\Omega_{DE}\sim 2/3$
while $\Omega_{matter}\sim 1/3$ and $\Omega_{tot}$ is very close to unity.
Inflation predicts $\Omega_{tot} = 1$, for all times after
inflation.
The matter contributions consist of (at least)
\begin{equation}
\Omega_{matter}=\Omega_{b}+\Omega_{\nu}+\Omega_{LSP} + \Omega_{a}  
\end{equation}
The most recent data is consistent with nucleosynthesis
estimates of $\Omega_{b}\sim 0.04$, where the baryons (b) in the universe
are mainly to be found in dark objects. 
The determined value of 
$\Omega_{CDM}\sim 0.3$ contains unknown relative contributions from
$\Omega_{LSP}$ and $\Omega_{a}$. 
Super-Kamiokande sets a lower limit on neutrino masses
$\sum_i m_{\nu_i}\geq 0.05$ eV which corresponds to $\Omega_{\nu}\geq 0.003$. 
In hierarchical neutrino mass models the lower bound is saturated
and interestingly the neutrino density is then comparable to
the visible baryon density $\Omega_{stars}\sim 0.005$.

\section{An Example}
\subsection{Why Intermediate Scale Inflation?}
We now wish to consider a specific example of a
model which addresses the particle
physics issues mentioned in the abstract, 
in order to illustrate many of the
general features that we have discussed above.
The brand of inflation most closely related to particle physics
seems to be hybrid inflation which may occur at a scale
well below the Planck scale, and hence be in the realm of particle
physics. The next question is what is the
relevant scale at which hybrid inflation takes place?
One obvious possibility is to
associate the scale of inflation with some grand unified theory
(GUT) symmetry breaking scale, as originally conceived by Guth. 
However it is somewhat ironic that 
hybrid inflation at the GUT scale faces the magnetic monopole
problem, which was precisely one of the original motivations
for considering inflation in the first place! Although in certain
cases this problem may be resolved, there are typically further symmetry
breaking scales below the GUT scale at which discrete symmetries
are broken, leading to problems with cosmological domain walls.

Intermediate scale hybrid inflation 
immediately solves
both the magnetic monopole problem and the domain wall problem.
The idea is simply that there is a period of hybrid inflation
occuring below the GUT scale at the PQ symmetry breaking scale
itself, in which the inflaton carries PQ charge and so the choice of
domain is fixed during inflation. The universe therefore inflates
inside a particular domain, and the magnetic monople relics
produced by the GUT scale symmetry breaking are inflated away.
This provides a powerful
motivation for intermediate scale inflation.

\subsection{Intermediate Scale Supersymmetric Inflation Model}
The model we consider is a variant of the 
NMSSM. This model has a SUGRA foundation
and leptogenesis and reheating has been studied
and preheating has been demonstrated
not to lead to over-production of either axions or gravitinos.
The model provides a solution to the strong CP problem and the
$\mu$ problem, with inflation directly solving the monopole
and domain wall problems at the inflation scale.
It is therefore a well motivated, successful model 
that has been well studied and does not appear to suffer
from any embarrassing problems, and is therefore a suitable
laboratory for our discussion here.
This variant of the NMSSM has the following superpotential terms involving the
standard Higgs doublets $H_{u},H_{d}$
and two gauge singlet fields $\phi$ (inflaton) 
and $N$,
\begin{equation}
 W = \lambda N H_{u} H_{d} - k \phi N^{2}  \label{eq:super}
\end{equation}
where $\lambda ,k$ are dimensionless coupling constants.
Notice that the standard NMSSM is recovered if we replace the inflaton $\phi$ 
by N.  However this leads to the familiar domain wall problems arising from 
the discrete $Z_{3}$ symmetry.  In this new variant, the $Z_{3}$
becomes a global $U(1)_{PQ}$ symmetry that is commonly 
invoked to solve the strong CP problem.
This symmetry is broken in the true 
vacuum by intermediate scale $\phi$ and N VEVs, 
where the axion is the pseudo-Goldstone
boson from the spontaneous symmetry breaking and constrains the size of the 
VEVs. With such large VEVs
this model should be regarded as giving an intermediate scale
solution to the $\mu$ problem, and as such will have collider signatures.

We can make the $\phi$-field real by a choice of the (approximately) massless 
axion field.  We will now regard $\phi$ and $N$ to be the real components of 
the complex singlets in what follows.
When we include soft SUSY breaking mass terms,
trilinear terms $A_{k}k \phi N^{2} + h.c.$ (for real $A_{k}$) and neglect the
$\lambda N H_{u} H_{d}$ superpotential term, we have the following 
potential:
\begin{equation}
 V=V_{0} + k^{2} N^{4} + \frac{1}{2} m^{2}(\phi) N^{2} 
  + \frac{1}{2} m_{\phi}^{2} \phi^{2}
     \label{eq:potential} 
\end{equation}
where $m^{2}(\phi)= m_{N}^{2} + 4k^{2} \phi^{2} - 2k A_{k} \phi $.
We can identify the various elements of the potential: $V_{0}$ arises
from some other sector of the theory, SUGRA for example, and dominates the
potential;
the soft SUSY breaking parameters $A_{k}$ and $m_{N}$ are
generated through some
gravity-mediated mechanism with a generic value of $O(TeV)$.

More details about how hybrid inflation may be implemented in this
model may be found in \cite{King:2001mt}. We only remark here
that it is non-trivial that a set of parameters exists that is
consistent with axion and SUSY physics and allows
the correct COBE perturbations to be achieved by radiative corrections.
Without SUSY one would be free to add soft scalar masses at will,
but with SUSY one must rely on the theory which either generates
soft masses of order a TeV, or sets them equal to zero as in no-scale
SUGRA, in which case
the radiative corrections, which are under control in the case
of SUSY predict the relevant value of the soft parameters, without
any further adjustable parameters. Thus SUSY is playing a crucial
role in the model which is why we refer to it as a Supersymmetric
Inflation Model.

\subsection{The Cosmological Constant Problem}
Notice that the SUGRA-derived potential contribution $V_{0}$ exactly cancels
with the other terms (by tuning) to provide agreement with the observed small
cosmological constant. 
Thus we assume that at the global minimum
$V(\langle \phi \rangle , \langle N \rangle )=0$ which implies that
$ V_0 = k^{2} \langle N \rangle^{4}$.
The height of the potential during inflation is therefore 
$V_{0}^{1/4} = k^{1/2} \langle N \rangle \sim 10^{8} GeV$.
Since the approach has a consistent way to set the large 
cosmological constant to zero, the absence of a real solution to this problem may not
be an obstacle to the implications of the approach.

\subsection{Parameter Counting and Singlino Mixing}
A relevant parameter count at this stage reveals two superpotential
effective parameters ($\lambda$ and $k$),
the two soft SUSY breaking parameters ($A_{k}$ and $m_{N}$),
plus the constant energy density $V_0$.
From these five parameters we have inflated the universe
with the correct COBE perturbations,
provided a $\mu$ term of the correct order of magnitude
and solved the strong CP problem.
They also govern the phenomenology of the singlet Higgs
and Higgsino components of $\phi$ and $N$ which may weakly mix
with the MSSM superfields $H_u,H_d$. 

The LSP will be the lightest eigenvalue of the full ``ino'' matrix,
extended in the usual way to include gaugino-higgsino mixing.
Clearly if $k<\lambda/2$ then a singlino will
be the LSP. In our case the singlino may be regarded as
a linear combination of axino and inflatino.

As usual in models based on an intermediate scale solution to the
$\mu$ problem the coupling of the 
singlino to the neutralinos means that $\tilde{S}$ nearly decouples.
However the conservation of R-parity means that eventually the
lightest neutralino produced in colliders must decay into the
singlino, and all the collider signatures discussed
in may apply. In the case that the lightest
neutralino leaves the detector before it decays into the singlino,
there will be no unconventional collider signature. In this case the
knowledge concerning a lighter singlino will
come from cosmology since the LSP relic density gets diluted
by the ratio of the singlino to lightest neutralino masses,
and direct dark matter searches will not see anything since the
singlino LSP will not scatter off nuclei.

\begin{table}[tbp]
\hfil
\begin{tabular}{cccccc}
\hline \hline 
 & $V_0$ & $k$  & $\lambda$ & ${\cal L}_{soft}$ &  ${\cal L}_{Yuk}$          
\\ \hline \hline
Inflation and COBE 
& $\surd$ & $\surd$  & -  &  $\surd$ & - \\
MSSM $\mu$ parameter 
& -      & - &  $\surd$ &  $\surd$ &   - \\
Fermion Masses, Mixings
& - &   - & -  & - & $\surd$ \\
SUSY collider physics
& - & $\surd$ & $\surd$ & $\surd$ & $\surd$ \\
Strong CP, axion abundance ($\Omega_a$)
& -  &    $\surd$ & $\surd$ & $\surd$ & - \\
Leptogenesis ($\Omega_b$)
& - & $\surd$ & $\surd$ & $\surd$ & $\surd$ \\
LSP abundance ($\Omega_{LSP}$)
& - & $\surd$ & $\surd$ & $\surd$ & - \\
\hline \hline
\end{tabular}
\hfil
\caption{\footnotesize 
This table illustrates the fact that a particular parameter of the
model (columns) simultaneously controls
several different aspects of particle physics
and cosmology (rows) which are thereby related.
${\cal L}_{soft}$ contains $A_k,m_N^2$ and the other soft parameters,
${\cal L}_{Yuk}$ contains the Yukawa coupling constants
controlling all fermion masses and mixing angles.}
\label{forbidden}
\end{table}

One of the main things emphasized in \cite{Kane:2001rb}
is the connection between the calculation
of relic densities and the other physics, via their common
parameters. This is summarised in Table 1 for the particular model
we have been discussing.
The same parameters that control the ino mass matrix
will also be involved in reheating of the universe after inflation, and
giving the relic densities of LSP and in leptogenesis
as we discussed in \cite{Kane:2001rb}. 
Different models may have different mechanisms to solve
some of the problems, different reheating and preheating, and so on, but will
still lead to a version of Table 1.

\section{How To Calculate the Size of the 
Dark Energy Density in Supersymmetric Inflation}

Is there anything that we can say about DE at the current time
from the perspective
of our supersymmetric particle physics based model of inflation?
Perhaps surprisingly the answer is positive: we shall show that 
we can deduce the present day value of dark energy density from the
model, together with the measured CMB temperature
and the Hubble constant, even if the model
does not yet specify the physics of dark energy \cite{Kane:2001rb}!

A key point of our approach is that a supersymmetric particle physics based
model of inflation enables us to calculate (in principle at least)
the (energy or number)
densities of all forms of radiation and matter 
(but excluding dark energy) at some
early time $t_{RH}$ after inflation and reheating has taken place,
corresponding to the start of the standard hot big bang.
For simplicity we consider only
one type of matter energy density $\rho_{matter}(t_{RH})$
(which may readily be obtained from the calculated number density) 
and radiation energy density $\rho_{\gamma}(t_{RH})$.
The argument may be
straightforwardly generalised to the case of several components
of radiation and matter.
Now, using the equations of the standard hot big bang,
we wish to obtain their values at the present time $t_0$, 
$\rho_{\gamma}(t_{0})$ and $\rho_{matter}(t_{0})$.
Without further information this is impossible since we need something to 
tell us when the present time $t_0$ is, and moreover the model does
not specify either $\rho_{DE}(t_{RH})$, or its equation of state,
both of which will influence the evolution of the universe.
Therefore let us input into our analysis the present day
observed CMB temperature $T_0=2.725{\rm K}$, which corresponds to
a photon density $\rho_{\gamma}(t_{0})=(2.115\times 10^{-4}{\rm eV})^4$,  
a photon number density $n_{\gamma}=410 {\rm cm^{-3}}$,
and, assuming three families of light neutrinos, an entropy density
$s=7.04n_{\gamma}$. 
Then, ignoring additional sources of entropy
between $t_{RH}$ and $t_0$ (such as electron-positron annihilation),
since we know the equations of state for photons and matter,
$\rho_{\gamma}\sim R^{-4}$ and $\rho_{matter}\sim R^{-3}$,
where $R$ is the scale factor of the universe, using the initial
values of $\rho_{\gamma}(t_{RH})$ and $\rho_{matter}(t_{RH})$
predicted by the model and the present value of 
$\rho_{\gamma}(t_{0})$ from observation, we find
$\rho_{matter}(t_{0})\approx \rho_{matter}(t_{RH})[\rho_{\gamma}(t_{0})/
\rho_{\gamma}(t_{RH})]^{3/4}$. We emphasise that this determination of
$\rho_{matter}(t_{0})$ is independent of the unknown dark energy.

Once we have obtained $\rho_{matter}(t_{0})$, 
from a combination of our model calculation and the
observed CMB temperature, as outlined above, we now use the observed
Hubble constant $H_0$, or equivalently the present day critical
density $\rho_{crit}$, to convert $\rho_{matter}(t_{0})$
into the various $\Omega_{matter}=\rho_{matter}(t_{0})/\rho_{crit}$.
Once $\Omega_{matter}$ is predicted
within some supersymmetric particle physics
based model of inflation, supplemented by measurements of the
CMB temperature and the Hubble constant,
then it is clear that
$\Omega_{DE}$ is also {\em predicted } to be 
\begin{equation}
\Omega_{DE} = 1- \Omega_{matter}
\label{lambda}
\end{equation}
Thus a model of inflation that is capable of predicting
$\Omega_{matter}$ using measurements of the
CMB temperature and the Hubble constant, is also capable of predicting
$\Omega_{DE}$ from Eq.\ref{lambda}.
This sum rule has been written down before,
including a curvature term and it has been discussed how to 
determine each of the components $\Omega_{DE}$ and $\Omega_{matter}$
from observation. 
What we are saying here is quite different from the empirical approach
to determining the components of this equation.
To begin with we are assuming inflation, so that the curvature
contribution is zero. Secondly we are only taking two inputs
from observation, namely the CMB temperature and the value
of the Hubble constant. Given these inputs we have shown how an
inflation model allows us to then {\em calculate} $\Omega_{matter}$,
and hence {\em deduce} $\Omega_{DE}$ from Eq.\ref{lambda}.

\section{Conclusion}
Over the next few years there will be considerable 
progress in cosmology from the Map and Planck explorer satellites, and 
in SUSY from the Tevatron and LHC. We believe the time is
ripe for a new closer synthesis of SUSY and inflation, and that
the most promising scenario will involve these theories
meeting at the intermediate scale. We have shown that
in this case one may hope to relate different phenomena
in cosmology and in particle physics in a much closer and
more predictive way than ever before. Finally
we have made the original observation that, given the
value of the CMB temperature and Hubble constant from observation,
an intermediate scale supersymmetric inflation model allows
the present day matter relic density to be calculated,
and hence the present day DE relic density to be determined
from Eq.\ref{lambda} even in the absence of any theory of DE.

\begin{center}
{\bf ACKNOWLEDGMENTS}
\end{center}
I would like to thank co-authors Mar Bastero-Gil, Gordy Kane and
David Rayner.  
I enjoyed conversations at COSMO-01 with many people especially
J. Cline, E. Copeland, J. Garcia-Bellido, J.E. Kim,
A. Kusenko, A. Liddle, A. Linde, C. Lineweaver, A. Mazumdar, 
C. Munoz, P. Nilles, K. Olive, J. Peacock, M. Pluemacher and S. Sarkar.

\end{document}